\documentclass[floats,twocolumn,pra,footinbib,floatfix]{revtex4}
\usepackage{eurosym}
\usepackage{graphics}
\usepackage[thmmarks]{ntheorem}
\usepackage{graphicx}
\usepackage{algorithm}
\usepackage{algpseudocode}
\usepackage{amsmath}
\usepackage{amsfonts}
\usepackage{amssymb}
\usepackage{float}
\usepackage{longtable}
\usepackage{flushend}
\usepackage{epsfig}
\usepackage{latexsym}
\usepackage{theorem}
\usepackage{bbm}
\usepackage{bm}
\usepackage{psfrag}
\usepackage{eurosym}
\usepackage[utf8]{inputenc}
\usepackage[T1]{fontenc}
\usepackage{physics}%
\usepackage{color}
\usepackage{multirow}
\usepackage{float}
\usepackage{hyperref}
\usepackage{mathrsfs}
\usepackage{cancel}
\usepackage{xcolor}
\usepackage{soul}
\DeclareMathOperator{\diag}{diag}

\begin{document}

\title{Continuous-Variable Quantum Key Distribution with Composable Security and Tight Error Correction Bound towards Constrained-Device Implementations}
\author{Panagiotis Papanastasiou$^{1,3}$}
\author{Carlo Ottaviani$^{2,3}$}
\author{Stefano Pirandola$^{2}$}
\author{Poonam Yadav$^{2}$}
\author{Marco Lucamarini$^{1,3}$} 

\affiliation{$^1$School of Physics, Engineering \& Technology, University of York, YO10 5FT York, U.K.}
\affiliation{$^2$Department of Computer Science, University of York, YO10 5GH York, U.K.}
\affiliation{$^3$York Centre for Quantum Technologies, University of York, YO10 5FT York, U.K.}

\begin{abstract} 
\noindent Constrained devices, such as smart sensors, wearable devices, and Internet of Things nodes, are increasingly prevalent in society and rely on secure communications to function properly. These devices often operate autonomously, exchanging sensitive data or commands over short distances, such as within a room, house, or warehouse. In this context, continuous-variable quantum key distribution (CV-QKD) offers the highest secure key rate and the greatest versatility for integration into existing infrastructure. A key challenge in this setting, where devices have limited storage and processing capacity, is obtaining a realistic and tight estimate of the CV-QKD secure key rate within a composable security framework, with error correction (EC) consuming most of the storage and computational power. To address this, we focus on low-density parity-check (LDPC) codes with non-binary alphabets, which optimise mutual information and are particularly suited for short-distance communications. 
We develop a security framework to derive finite-size secret keys near the optimal EC leakage limit and model the related memory requirements for the encoding process in one-way error correction. This analysis facilitates the practical deployment of CV-QKD, particularly in constrained devices with limited storage and computational resources.
\end{abstract}
\maketitle

\section{Introduction}
\noindent Quantum key distribution (QKD)~\cite{BB84} allows two parties to establish a common secret key, which can later be used in symmetric cryptographic primitives. Its security relies on the fundamental principles of quantum physics rather than computational complexity conjectures~\cite{rev1,rev2,rev3,UsenkoRev} and constitutes, along with post-quantum cryptography~\cite{post-quantum_rev}, the leading candidate for countering quantum threats, such as an eavesdropper equipped with a quantum computer.

Initially, QKD was developed for discrete-variable (DV) systems, which use discrete degrees of freedom, such as the polarization of the electromagnetic field. Later, protocols based on continuous degrees of freedom, such as the quadratures of the electromagnetic field, i.e., continuous-variable (CV) systems~\cite{gaussianQI}, emerged, offering high performance in the asymptotic regime and over short distances, as well as compatibility with existing technological infrastructure. Recent studies have advanced both the security~\cite{lev_definetti,ImprovedRates} and the experimental performance~\cite{exp_1,exp_2} of CV-QKD, bringing it close to the repeaterless PLOB bound~\cite{PLOB} and making it comparable to DV-QKD.

The most common and earliest CV-QKD protocols employ Gaussian modulation of coherent states (GMCS), utilising homodyne detection~\cite{GG2002} or heterodyne detection~\cite{WR2004} in direct or reverse reconciliation (RR)~\cite{RRGMCS}. CV-QKD has also been extended to protocols using different frequencies (thermal states), two-way communication~\cite{thermal-sates,two-way,two-way thermal,thermal-fnsz,teraherz}, and network settings~\cite{CV-MDI0,CV-MDI1,CV-MDI2,cv-mdi-fnsz1,cv-mdi-fnsz2,CVMDICOMP1,CVMDICOMP2,pp_sim_three,CVMDIrev,conferencing,fdchannel,end-to-end}. Schemes employing post-selection~\cite{ps_weedbrook,post_selection_Kieran,Gpost_selection} and discrete modulation~\cite{DA_0,DA_panos_1,DA_1,DA_panos_2,DA_2,DA_3} have also been considered.

Recent demonstrations of continuous-variable quantum key distribution (CV-QKD)~\cite{HajomerI, HajomerII, Aldama, Pietri, LangLi} have used integrated photonic systems~\cite{Silverstone}, paving the way for deploying QKD protocols on constrained devices such as drones, vehicles, IoT (Internet of Things) devices, and wearables. As these devices become ubiquitous in industrial, healthcare, and consumer networks, it is nearly impossible at design time to predict which endpoints will eventually handle mission-critical or privacy-sensitive data~\cite{Roman2013_IoTSec}. Moreover, untrusted edge nodes often represent the weakest link in a defence-in-depth architecture, potentially creating backdoors into deeper network layers~\cite{Neshenko2019_IoTSurvey}.

Post-quantum cryptography (PQC)~\cite{LiuMoody2024_PQCReview} remains the most mature “plug-and-play” solution for providing quantum-resistant security and is already under benchmarking for constrained-device scenarios~\cite{Carlo_CD}. It requires a smooth integration of standardised algorithms into existing stacks~\cite{Joseph2022_PQCPerspective}. However, it still relies on unproven hardness assumptions and cannot match the information-theoretic security (ITS) of QKD~\cite{Mosca2018_Cybersecurity,rev3}.

Recognising this trade-off, one can view cryptographic security solutions as spanning a spectrum from classical cryptography through PQC to QKD, with each step requiring greater implementation effort but delivering stronger secrecy. Rather than relying on any single approach, emerging best practices advocate hybrid PQC and QKD schemes that combine the strengths of both methods~\cite{Garms2024_Hybrid,Fedorov2023_HybridInfra,Buruaga2025_MACsec}. Therefore, beyond the challenges imposed by their small form factor, it is essential to identify the data post-processing limitations of such devices to predict the implementation performance of CV-QKD protocols given a specific level of security.

In this study, we focus on the GMCS protocol, which is particularly effective in short- to moderate-loss regimes, and particularly quantify the post-processing requirements, especially error correction (EC), which can significantly impact the protocol's performance~\cite{practical_ec_1,practical_ec_2,practical_ec_3,practical_ec_4,oneway_asymptotic, ma_sim_one, ma_sim_two, RealTimeThesis}. Inspired by the analysis of Ref.~\cite{ImprovedRates}, which imposes strict bounds on the secret key length, we quantify the protocol's performance under composable security with finite-size effects.

Furthermore, we adapt the tight one-way EC bound from Ref.~\cite{TomaLeak} to non-binary low-density parity-check (LDPC) codes~\cite{ldpc1,ldpc2,Multi-Repeat LDPC} to compute secret key rates with near-optimal performance. This allows us to estimate the storage requirements of the encoder during EC. Although this represents only half of the complete EC procedure, it provides valuable insight in scenarios where there is a strong asymmetry in computational resources between the CV-QKD transmitter and receiver. Specifically, the transmitter, which performs the encoding, can be lightweight and agile, while the receiver, responsible for decoding, may be bulkier and better suited for computationally intensive tasks.

This scenario arises, for example, in networks of small sensors transmitting to a central processing unit, as seen in smart home sensors, wearable devices,  IoT systems~\cite{IoT_0,IoT_Carlo,IoT_panos,Iot_1,IoT_2,IoT_3}, drones~\cite{drones_QKD} or free space implementations~\cite{RazaviIndoors,free-space}. These devices play a crucial role in modern society, supporting the technological infrastructure by enabling automation, real-time monitoring, and critical decision making. As they become increasingly integrated into daily life, cyber threats targeting them pose significant risks to public safety. We suggest feasible cases where CV-QKD can be used to safeguard against these threats.

In Sec.~\ref{skl}, we introduce the composable secret key length bound. In Sec.~\ref{sec:leakage}, we derive the information leakage in one-way EC for non-binary LDPC codes. In Sec.~\ref{main_res_tight_bound}, we present the final formula for the secret key rate, and in Sec.~\ref{LDPC_CODES}, we introduce one-way EC with non-binary LDPC codes. Finally, we link these results to an estimate of the storage required for the encoding of the EC procedure to achieve near-optimal performance. Sec.~\ref{Results} presents our numerical results, and Sec.~\ref{conc} concludes our work.

\section{Composable security rate with tight leakage bound \label{fnsz}}
To provide an accurate and richer description of the security aspects of GMCS protocols, our analysis takes into consideration finite-size effects along with data post-processing contributions, which are also relevant to assess the implementation and various performance trade-offs. We adopt the security proof of Ref.~\cite{ImprovedRates} that contains parameters such as the failure probability of EC, the reconciliation efficiency or the residual probability of EC which are connected to the EC leakage needed by the parties to reconcile their raw key. 

These parameters can be calculated  by applying  an EC procedure after running simulations~\cite{ma_sim_one,ma_sim_two}. predicting accurate results for the performance of the system. However, theoretical works  can predict the performance of the EC procedure assuming specific techniques. In particular, Ref.~\cite{TomaLeak},  provides bounds for the near-optimal (but achievable) performance of EC based on (binary) LDPC codes assuming finite-size effects.

In this section, we extend the previous analysis to the non-binary LDPC codes connecting the EC performance and security parameters of the secret key rate formula from Ref.~\cite{ImprovedRates}, a useful tool for benchmarking the behaviour of the CV-QKD protocols in different settings. As we will discuss in the next section, we can connect this result to the storage requirements for EC encoding.

\subsection{Secret key length with composable terms\label{skl}}
Before identifying the relevant parameters for EC performance connected to the non-binary LDPC codes, we present and summarize results from Ref.~\cite{ImprovedRates} about the secret key rate in a composable framework.

The secret-key length is upper-bounded by the following expression (see also Ref.~\cite[Eq.~(36)]{ImprovedRates}) 
\begin{equation}\label{main_comp_bound_zero}
s^{\epsilon_\text{cor}+\epsilon_\text{h}+\epsilon_\text{s}}_n\leq n R_\infty-\sqrt{n}\Delta^{\epsilon_\text{s}}_\text{aep}(h)+\theta,
\end{equation}
when one considers collective Gaussian attacks assuming that the protocol did not abort, the extracted key is correct, with probability larger than $1-\epsilon_\text{cor}$, and secret with probability larger than $1-\epsilon_\text{sec}$. The parameter $\epsilon_\text{sec}=\epsilon_\text{s}+\epsilon_\text{h}$, where $\epsilon_\text{s}$ is the smoothing parameter and $\epsilon_\text{h}$ is the probability of failure of Privacy Amplification (PA).
The asymptotic key-rate $R_{\infty}$ in Eq.~(\ref{main_comp_bound_zero}) is given by
\begin{equation}\label{rate-asy}
R_\infty=H(\mathsf{k})-\chi(\mathsf{k}:E)-\text{leak}_\text{ec},
\end{equation}
where the secret key variable $\mathsf{k}$ takes values from the alphabet $\mathcal{K}=\{0,1,\dots,2^{hd}-1\}$ for $d$-bit digitization of the normalized quadrature results~\cite{ma_sim_one,ma_sim_two,ImprovedRates}. The quantity $\Delta^{\epsilon_\text{s}}_\text{aep}$, given by the expression 
\begin{equation}\label{delta_aep}
\Delta^{\epsilon_\text{s}}_\text{aep}(h)\simeq 4\log_2\left(\sqrt{2^{hd}}+2\right)\sqrt{\log_2(2/\epsilon_\text{s}^2)},
\end{equation}
accounts for the penalty due to finite-size effects, while
\begin{equation}\label{theta}
\theta=\log_2(2\epsilon_\text{h}^2\epsilon_\text{cor})
\end{equation}
is the penalty paid for non-ideal verification and Privacy Amplification while $n$ is the block size of the raw key after channel parameter estimation (PE) and $\text{leak}_\text{ec}$ is the error correction (EC) leakage.

The quantity $H(\mathsf{k})$, in Eq.~(\ref{rate-asy}), is the Shannon entropy of the key variable $\mathsf{k}$ and $\chi(\mathsf{k}:E)$ is Eve’s Holevo information. In this description, the dummy variable $h=\{1,2\}$ distinguishes between homodyne ($h=1$) and heterodyne detection ($h=2$). More specifically, for heterodyne protocol, the digitized outcomes are in the vectorial form $(k_q,k_p)$, that can be concatenated as
\begin{equation}
\mathsf{k}=k_q 2^d+k_p,
\end{equation}
without loss of any information because the mapping $(k_q,k_p)\leftrightarrow \mathsf{k}$ is one-to-one, and with Shannon entropies related by the following mathematical expression $H(\mathsf{k})=2H(k)$ (see Appendix ~\ref{VC} for details).
For error correction the parties may now decide to use the vectorial form $\{k^{(1)}_{q},k^{(1)}_{p},k^{(2)}_{q},k^{(2)}_{p},\dots,k^{(n)}_{q},k^{(n)}_{p}\}$.

From here, we apply the procedure given in \cite[Eq.~(48)]{ImprovedRates}, to write Eq.~\eqref{rate-asy} as 
\begin{equation}
R_\infty=hH(k)-\chi(x:E)-\text{leak}_\text{ec},
\end{equation}
where variable $x$ is the continuous-variable version of $\mathsf{k}$. Note that for the homodyne protocol $k$ and $\mathsf{k}$ are equivalent forms of the key variable.
The steps to obtain the Holevo bound for the reverse reconciliation (RR) protocol have been detailed in Ref.~\cite{ImprovedRates}, hence we give only the final expression for the covariance matrix (CM) for the direct reconciliation (DR) protocol in Appendix~\ref{sum_SKR}.

We can now calculate the secret-key rate after channel PE, replacing in the Holevo bound the channel parameters, excess noise $\xi$ and transmissivity $\tau$ with their worse-case scenario values $\tau^{\epsilon_\text{pe}}$ and $\xi^{\epsilon_\text{pe}}$~\footnote{In a QKD session, we compute
the estimators and their variances directly from the experimental data. These are those
effectively used for the worst case estimators in a practical implementation. For theoretical analysis, one employs asymptotic or large-block approximations. Since in this analysis some of the estimators are a sample mean of independent observations, the Central Limit Theorem ensures approximate Gaussianity, and the Berry–Esseen theorem provides a quantitative bound on the deviation from the Gaussian limit which for the parameters used in this study (see Table~\ref{table:comp_rate_params}) and block sizes $2.5 \times 10^{4}-10^6$ is less than $10^{-2}-10^{-3}$. Although this bound is conservative, results based on Stein’s method and modern Gaussian-approximation theory show that practical convergence is typically faster than the worst-case rate suggests~\cite{ChenShao,Chernozhukov,Tomamichel}. This supports the Gaussian modelling adopted in our numerical worst-case analysis.}, respectively, given in \cite[Eqs.~(85) and~(86)]{ImprovedRates}. These values are dependent on the efficiency $\eta_d$, the electronic noise $u_\text{el}$ of the detection, and the number of signals. We then obtain the expression below
\begin{equation}\label{asym_PE_rate}
R^{\epsilon_\text{pe}}_n=hH(k)-\text{leak}_\text{ec}-\chi(x:E)|_{\tau^{\epsilon_\text{pe}},\xi^{\epsilon_\text{pe}}},
\end{equation}
encapsulating the finite-size dependence of the secret key rate due to PE and replaces $R_\infty$ in Eq.~\eqref{main_comp_bound_zero}.

\subsection{Theoretical estimation of EC leakage\label{sec:leakage}}
In the following discussion, we bound the leakage term of the previous formula adopting the approach of Ref.~\cite{TomaLeak} after considering the non-binary LDPC case. Then we connect it with the EC performance and security parameters such as the reconciliation efficiency, successful verification probability and EC failure probability.

In case of one-way reconciliation, where low-density parity-check (LDPC) codes~\cite{ldpc1,ldpc2} are used (see also Sec.~\ref{Regular LDPC}), the EC leakage term can be upper-bounded by the number of syndrome bits, given by
\begin{equation}\label{leak_ub}
\text{leak}_\text{ec}\leq n^{-1}\log_2|\mathcal{M}|,
\end{equation}
where $\mathcal{M}$ is the alphabet of the syndrome strings. One may calculate the size of the alphabet $|\mathcal{M}|$ via EC simulations, as done in Ref.~\cite{oneway_asymptotic} for asymptotic security analysis and in Refs.~\cite{pp_sim_three,ma_sim_one,ma_sim_two} for composable security analysis. 

However, in a complete theoretical analysis, one may use the asymptotic bound (Slepian-Wolf coding~\cite{SWC}) stating that
\begin{equation}\label{slwlf}
\log_2|\mathcal{M}|- nh H(k|y)\geq 0,
\end{equation}
where $H(k|y)$ is the conditional Shannon entropy of $k$ conditioned on the continuous variable $y$ of the other party. In fact, when considering finite-size effects one can use a more rigorous bound~\cite{TomaLeak} providing a tighter estimate of the performance of information reconciliation. Such a bound is given by

\begin{equation}\label{tight_bound}
\abs{\log_2|\mathcal{M}|-n hH(k|y)-\sqrt{n}\Delta^{\epsilon_\text{ec}}_\text{leak}(h)}\leq \delta(n),
\end{equation}
where
\begin{equation}\label{delta_leak}
\Delta^{\epsilon_\text{ec}}_\text{leak}(h)=\sqrt{hV(k|y)}\Phi^{-1}(1-\epsilon_\text{ec})
\end{equation}
with $\Phi$ being the cumulative normal distribution and $\epsilon_\text{ec}$ is a bound for $P[\hat{k}^{hn}\neq k^{hn}]$ and  $\hat{k}^{hn}$, $k^{hn}$  are the strings of the parties after EC. The right-hand side of Eq.~\eqref{tight_bound} is  given by 
\begin{equation}
\delta(n)=\frac{1}{2}\log_2hn+\mathcal{O}(1)
\end{equation}
while the conditional entropy and the conditional entropy variance are given by 
\begin{equation}
H(k|y)=\mathbb{E}\left[-\log_2p(k|y)\right],~V(k|y)=\text{Var}\left[-\log_2p(k|y)\right].
\end{equation}

We now extend the calculation of Eq.~\eqref{tight_bound} to non-binary alphabets (i.e., $d>1$) using the expression of $p(k|y)$ given in Eq.~(68) of Ref.\cite{ma_sim_one}. That allows the parties to achieve higher mutual information at short distances, because when $d$ increases, the entropy of digitized variables approach that of their continuous-variable counterpart. 
 
To minimize the probability of errors in the final key string, the parties apply the verification step: one party sends the syndrome and a hash of the raw key $k^{hn}$ with collision probability $\epsilon_\text{cor}$. The other party will compare this with the hash of the guessed string $\hat{k}^{hn}$ and, if they match, then the protocol can continue with probability $p_\text{ec}$ (we denote this event with $\top$), otherwise they will abort ($\bot$). In this case, the conditional probability of error is bounded by $P[\hat{k}^{hn}\neq k^{hn}|\top]\leq\epsilon_\text{cor}/p_\text{ec}$ which abides with the security definition in Refs.~\cite{Portmann, ImprovedRates} about the joint probability
\begin{equation}
P[\hat{k}^{hn}\neq k^{hn} \wedge \top]=p_\text{ec}P[\hat{k}^{hn}\neq k^{hn}|\top]\leq \epsilon_\text{cor}.
\end{equation}
To be more specific, in an honest implementation for Eve the parameter estimation step completely defines and bounds the channel parameters $\tau$ and $\xi$ prior to error correction. Based on these estimations, the parties decide the probability of successful verification $p_\text{ec}$ along with the collision probability $\epsilon_\text{cor}$. Through them they can decide the largest probability of failure that their EC LDPC scheme can tolerate given the previous parameters. Therefore, we need to calculate an upper bound for the marginal probability connected to Eq.~\eqref{delta_leak}
\begin{align}
P[ \hat{k}^{hn} \neq k^{hn}]=&P[ \hat{k}^{hn} \neq k^{hn}\wedge \top]+P[ \hat{k}^{hn} \neq k^{hn}\wedge \bot]\notag\\
\leq &\epsilon_\text{cor}+(1-p_\text{ec})P[\hat{k}^{hn} \neq k^{hn}|\bot]\notag\\
\leq&\epsilon_\text{cor}+(1-p_\text{ec})\notag\\
=&1-(p_\text{ec}-\epsilon_\text{cor}):=\epsilon_\text{ec}.
\end{align}

Therefore, from Eq.~\eqref{tight_bound}, we have that
\begin{equation}\label{leakage}
\log_2|\mathcal{M}|\leq n hH(k|y)+\sqrt{n}\Delta^{\epsilon_\text{ec}}_\text{leak}(h)+\delta(n)
\end{equation}
to be used in Eq.~\eqref{leak_ub} and consecutively in ~\eqref{asym_PE_rate}. We then obtain
\begin{align}
&R^{\epsilon_\text{pe}+\epsilon_\text{ec}}_n=hH(k)-hH(k|y)-\sqrt{n^{-1}}\Delta^{\epsilon_\text{ec}}_\text{leak}(h)-\frac{\delta(n)}{n}\notag\\
&~~~~~~~~~~~~~~~~~~~~~~~~~~~~~~~~~~~~~~~~-\chi(x:E)|_{\tau^{\epsilon_\text{pe}},\xi^{\epsilon_\text{pe}}}\\
&=h I(k:y)
-\chi(x:E)_{\tau^{\epsilon_\text{pe}}, \xi^{\epsilon_\text{pe}}}-\sqrt{n^{-1}}\Delta^{\epsilon_\text{ec}}_\text{leak}(h),
\label{asym_PE&EC_rate}
\end{align}
where we  omitted terms $\mathcal{O}\left(\frac{\log_2 n}{n}\right)$ in our calculations and $I(k:y)=H(k)-H(k|y)$ is the mutual information between $k$ and $y$. This is the rate that includes finite-size effects  connected only to PE and the performance of EC.

We also can group the mutual information and $\Delta^{\epsilon_\text{ec}}_\text{leak}$ term in Eq.~\eqref{asym_PE&EC_rate}, to define the quantities
\begin{equation}\label{zeta-leak}
\zeta_\text{leak}hI(k:y):=hI(k:y)-\Delta^{\epsilon_\text{ec}}_\text{leak}(h)/\sqrt{n}
\end{equation}
and
\begin{equation}\label{zeta-digit}
\zeta_\text{digit}:=hI(k:y)/I\left(x:y\right),
\end{equation}
where the mutual information $I(x:y)$ refers to the Gaussian variables as described in~\cite[Eq.~(83)]{ImprovedRates}. Using Eq.~\eqref{zeta-leak} and \eqref{zeta-digit} into Eq.~\eqref{asym_PE&EC_rate}, we can rewrite the rate as follows
\begin{align}\label{rate_zeta}
R^{\epsilon_\text{pe}+\epsilon_\text{ec}}_n&=\zeta I\left(x:y\right)-\chi(x:E)|_{\tau^{\epsilon_\text{pe}}, \xi^{\epsilon_\text{pe}}},
\end{align}
where
\begin{equation}\label{rec_eff_theo}
\zeta=\zeta_\text{digit}\zeta_{\text{leak}}.
\end{equation}

\subsection{Secret key rate with tight estimation of EC performance\label{main_res_tight_bound}}
Before we make the connection with the storage requirements for EC encoding, we present the main result of our previous discussion (see also Appendix~\ref{sum_SKR}).

Replacing $R^{\epsilon_\text{pe}}_n$ of Eq.~\eqref{asym_PE_rate} with $R^{\epsilon_\text{pe}+\epsilon_\text{ec}}_n$ from either Eq.~\eqref{asym_PE&EC_rate} or~\eqref{rate_zeta}, we obtain
\begin{equation}\label{secret_key_ration}
s^\epsilon_n/n \leq r^\epsilon_n:=\\
R^{\epsilon_\text{pe}+\epsilon_\text{ec}}_n
-\frac{\Delta^{\epsilon_\text{s}}_\text{aep}(h)}{\sqrt{n}}+\frac{\theta}{n}  -\mathcal{O}(\log_2(n)/n).
\end{equation}
This gives the highest number of bits per signal that can be extracted with security $\epsilon$ and a tight estimation of EC leakage, and it can be used to compute the final (expected) secret key rate,
\begin{equation}\label{main result}
R:=p_\text{ec}(n/N)r^\epsilon_n,
\end{equation}
where $N$ is the number of the total signals in the block.
Here, the choice of $p_\text{ec}$ entails a trade-off: a higher target $p_\text{ec}$ requires higher leakage and as a result a smaller key.

\section{LDPC codes for non-binary variables\label{Regular LDPC} and storage requirements\label{LDPC_CODES}}

Here, we describe the EC encoding procedure with LDPC codes connecting it to its storage requirements. We further use the bound of the previous section to quantify numerically the leakage and the storage needed for a specific secret key rate performance. We consider an EC scheme, where the encoder obtains a string $k^{hn}$ and computes the syndrome $\mathbf{H}k^{hn}=s^r$ where $\mathbf{H}$ is a $r \times hn$ parity-check matrix. In general, every element of the matrix belongs to the Galois field $k\in\mathcal{GF}=\{0,1,\dots,2^d-1\}$.  More specifically, the matrix is a representation of a Tanner graph~\cite{ma_sim_one} with $hn$ message nodes and $r$ parity -check nodes. When a message $i=1,\dots,hn$ is included in a parity-check $j=1,\dots,r$, there is an edge between the corresponding message node and parity-check node while the entry $H_{ji}$ of the associated parity-check matrix in this intersection is chosen randomly from $k=1,\dots,2^d-1$. 

One then may calculate the corresponding code rate
\begin{equation}\label{code_rate}
R_\text{code}=\frac{hn-r}{hn}=1- R_\text{synd}
\end{equation}
with $R_\text{synd}$ being the syndrome rate. The code is usually designed by assuming that each message participates in $d_\text{v}$ (column weight) checks and every check contains $d_c$ (row weight) messages. 
Then the number of edges must follow $hn d_\text{v}=r d_\text{c}$ (for sparse matrices, i.e., $d_\text{v}< d_\text{c}\ll r < hn$). Therefore, by replacing $r:= hn\frac{d_\text{v}}{d_\text{c}}$ in Eq.~\eqref{code_rate}, we obtain another equivalent expression 
\begin{equation}\label{Rdes}
R_\text{code}=1-\frac{d_\text{v}}{d_\text{c}},
\end{equation}
where we usually set $d_\text{v}=2$ as it gives the best performance in decoding~\cite{ma_sim_one}. This means that the rates for regular LDPCs are given by $$R_\text{code}=0.333,0.5,0.6,0.666,0.714,0.75,\dots,0.777,0.8\dots.$$

On the other hand, irregular LDPC codes can be designed where the column and row weights are not constant. Through the probability distributions of the weights, one defines their means $\bar{d}_\text{c}$ and $\bar{d}_\text{v}$, respectively. Therefore, we have
\begin{equation}
R_\text{code}=1-\frac{\bar{d}_\text{v}}{\bar{d}_\text{c}},
\end{equation}
which allows for more flexible values than before. Note that Eq.~\eqref{Rdes} allows us to generalize the notion of $R_\text{code}$ to the irregular code case. However, the performance of these codes is not particularly stable, i.e., different structures with the same $R_\text{code}$ behave very differently in terms of correcting different levels of signal-to-noise ratio (SNR) or in terms of probability of successful EC.
Then from Eq.~\eqref{leak_ub}, we have that
\begin{equation}
\text{leak}_\text{ec}\leq n^{-1} \log_2|\mathcal{M}|=n^{-1}dr=hdR_\text{synd}.
\end{equation}

One then may define a tight approximation for the  optimal $R_\text{synd}$ by  Eq.~\eqref{leakage}, where
\begin{equation}\label{rstarsynd}
R^\ast_\text{synd}=H(k|y)/d+\Delta^{\epsilon_\text{ec}}_\text{leak}(h)/(dh\sqrt{n})+\delta(n)/(dhn).
\end{equation}
Then, one may find values for the $R_\text{code}$ that perform closely to the previous tight bound in terms of leakage (the corresponding structure of the codes needs to be optimized to achieve a certain probability of successful EC for a specific SNR). These are given through Eq.~\eqref{code_rate} by
\begin{equation}\label{Rcode_ast}
R^\ast_\text{code}=1-R^\ast_\text{synd},
\end{equation}
where by $n\rightarrow\infty$ we arrive at the asymptotic expression $$R^\ast_\text{code}|_{n\rightarrow\infty}=1-d^{-1}H(k|y).$$

Then, we can calculate the required memory to store the parity-check matrix $\mathbf{H}$ as
\begin{align}
M_\text{code}:=hn \times  \log_2|\mathcal{M}|=hn dr=n^2 h ^2 d R_\text{synd} \label{mem_gen}.
\end{align}
For example, for a protocol using homodyne detection, with a block size of the raw key equal to $n=10^5$ and $d R^\ast_\text{synd}=4 \times 0.667$ we obtain the parity-check matrix storage to be around $3.34$ GBs while for a protocol using heterodyne detection it will be 4 times larger.

In the compressed row storage (CRS) format~\cite[Sec.~6.3]{RealTimeThesis}, one will need $\bar{d}_\text{v} \times h n \times d$ bits for storing the non-zero elements of the parity-check matrix, $ \bar{d}_\text{v} \times h n \times\lceil \log_2(h n) \rceil$ bits for storing the column indices, and $ \left(\frac{\log_2|\mathcal{M}|}{d}+1\right)\times\left\lceil\log_2\left(\bar{d}_\text{v} h n\right)\right\rceil $ bits for the row pointers. Gathering all these terms, one may calculate the sparse matrix representation storage as
\begin{align}
M_\text{sparse}=&\bar{d}_\text{v}hnd+ \bar{d}_\text{v }hn\lceil \log_2(h n) \rceil\notag\\&+\left(hnR_\text{synd}+1\right)\left\lceil\log_2\left(\bar{d}_\text{v} h n\right)\right\rceil.\label{mem_gen_sparse}
\end{align}

For a protocol using homodyne detection with $\bar{d}_\text{v}=2$, $n R_\text{synd}=0.667\times 10^5$, $d=4$, and $n=10^5$, we have that the sparse matrix storage will be approximately equal to $0.67$ MB while for a protocol using heterodyne detection $2$ times larger.
\begin{table}[t]
\centering
\begin{tabular}{|c|c|c|c|c|c|}
\hline
$N (10^5)$ & $n (10^5)$ & $R^{\ast}_\text{code}$ & $M^{\ast}_\text{sparse}$(MB) & $R_\text{code}$& $M_\text{sparse}$(MB) \\
\hline
 $ 1$         & $0.676$    & $0.78$  & $0.389$ & $0.777$ & $0.464$ \\
\hline
 $2 $ & $1.6 $    & $0.78$    & $0.95$ & $0.777$& $1.1$ \\
\hline
 $4$ & $3.2 $ & $0.7949$       & $2$  & $0.8$   & $2.3$    \\
\hline
\end{tabular}
\begin{centering}
\caption{We have chosen block sizes $1-4 \times 10^5$ and found the associated $R^{\ast}_\text{code}$ and $M^{\ast}_\text{sparse}$ from Fig.~\ref{fig:script_db=0.02_HomDR_data}. Then with the associated parameters, we have created parity-check matrices in the CRS format in Python with $R_\text{code}\approx R^{\ast}_\text{code}$, using regular non-binary LDPC codes. We finally calculated the actual storage needed for these cases and listed the results under the  $M_\text{sparse}$ column. These points have been depicted in the bottom panel of Fig.~\ref{fig:script_db=0.02_HomDR_data} with red ink following the predicted performance.\label{table:VALUE_TABLE}}
\end{centering}
\end{table}

Note here that Eqs.~\eqref{mem_gen} and~\eqref{mem_gen_sparse} calculate the practical storage associated with the parity-check matrix of a code with rate $R_\text{code}=1-R_\text{synd}$. Theoretically, one may compute tight bounds for these quantities through Eq.~\eqref{tight_bound} and obtain an approximate prediction for them. Thus, we may write
\begin{align}
M^\ast_\text{code}&=(h n)^2 d R^\ast_\text{synd}\label{mem_gen_theo}\\
M^\ast_\text{sparse}&=\bar{d}_\text{v}hn(d+\left\lceil\log_2(hn)\right\rceil)\notag\\
+&\left(hnR^\ast_\text{synd}+1\right)\left\lceil\log_2\left(\bar{d}_\text{v}hn\right)\right\rceil\label{mem_gen_sparse_theo}.
\end{align}
The most accurate estimation for these quantities is to simulate the results, i.e., create parity-check matrices for different block sizes (see Table~\ref{table:VALUE_TABLE}) with similar parameters and store them in CRS format. We have done this using the EC encoding script 
developed in the simulation library for the GMCS protocol with heterodyne detection~\cite{softquanta2021}. This may give different results due to a particular choice of the type of variables in the script or other software parameters used to describe the parity-check matrix in the specific format that may add an overhead, which is not included in Eq.~\eqref{mem_gen_sparse_theo}. Although this formula cannot give very accurate results, it is quite simple and can provide key insights for the encoding function. Notice that Eq.~\eqref{mem_gen_sparse_theo} through Eq.~\eqref{rstarsynd} associates the secret key rate performance (see Eq.~\eqref{main result}) to the required storage when both are calculated using the same parameter values.

\begin{table}[t]
\centering
\begin{tabular}{|c|c|c|}
\hline
 $\xi$&  excess noise &0.01 \\
\hline
 $\eta_d$& detection efficiency  &0.8 \\
\hline
 $ u_\text{el}$ & electronic noise& 0.01 \\
\hline
 $\tau $ & channel transmissivity &$10^{-\text{dB}/10} $ \\
\hline
$\epsilon_\text{h} $  &failure of privacy amplification&  $2^{-32}$  \\
\hline
 $\epsilon_\text{cor} $  & correctness  &$2^{-32}$   \\
\hline
$\epsilon_\text{pe} $  &  failure of parameter estimation &  $2^{-32}$  \\
\hline
$\epsilon_\text{s} $  &smoothing parameter   &$2^{-32}$   \\
\hline
\end{tabular}
\begin{centering}
\caption{Here we present the common parameters used to plot the secret key rate of Eqs.~\eqref{main result} in all figures.\label{table:comp_rate_params}}
\end{centering}
\end{table}

\begin{figure}[t]
\centering
\includegraphics[width=0.4\textwidth]{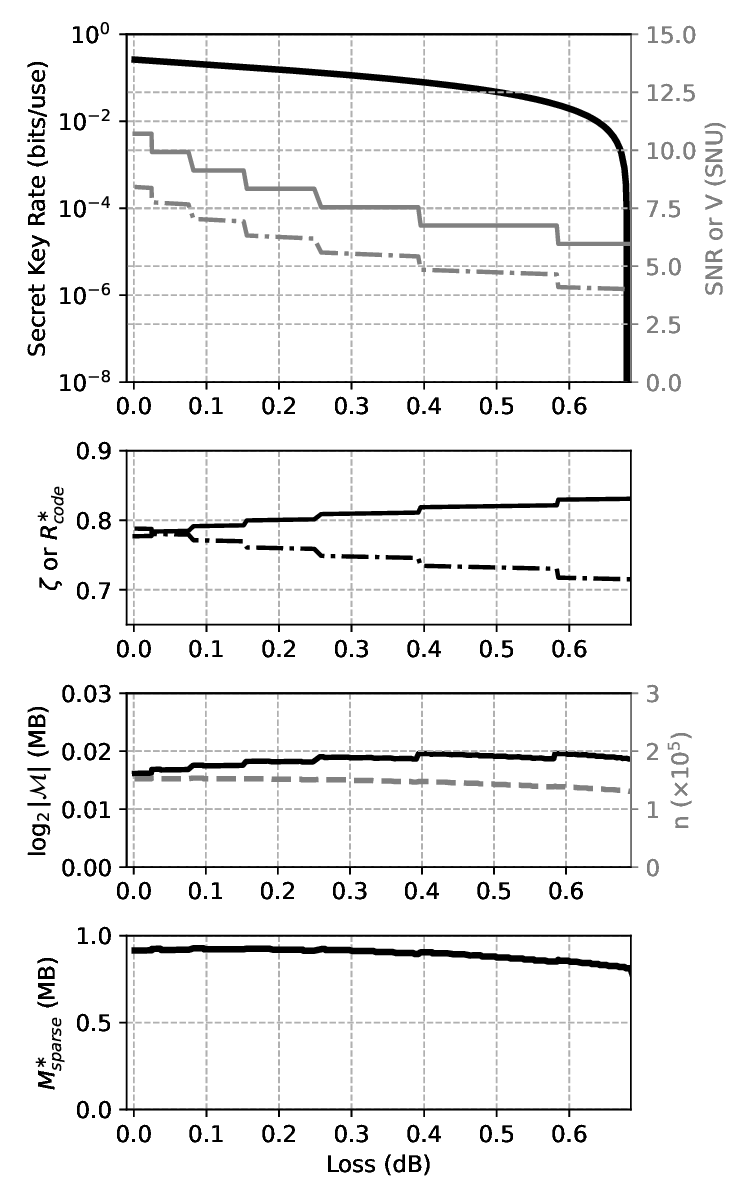}
\caption{In the top panel, we plot the secret key rate of Eq.~\eqref{main result} (black line) of the Gaussian modulation protocol with coherent states and homodyne detection in direct reconciliation against the loss in dB. We have optimized the secret key rate over the Gaussian modulation variance $V$ (grey thin line) and the PE ratio of sacrificed channel uses. We have assumed $N=2\times 10^5$,  $d=4$, and $p_\text{ec}=0.9$. The rest of the parameters are given in Table~\ref{table:comp_rate_params}. We plot also the corresponding SNR (gray dashed-dotted line) by replacing the optimal $V$ in the formula given by~\cite[Eq.~(33)]{ma_sim_two}. In the second panel, we plot the corresponding values of the reconciliation efficiency $\zeta$ (black line), which follow the same pattern as the Gaussian modulation variance values. However, we can see that for a constant value of $V$ the reconciliation efficiency increases linearly with the loss in dB. Subsequently, we plot the associated $R^{\ast}_\text{code}$ (black dash-dotted line). In the third panel, we plot the corresponding block size (gray dashed line) after optimizing the number of sacrificed channel uses during PE. We plot also the associated leakage (black line), i.e., $\log_2|\mathcal{M}|$ from Eq.~\eqref{leakage}. In the last panel, we plot Eq.~\eqref{mem_gen_sparse_theo} for the corresponding values of the secret key rate and loss in dB.}
\label{fig:script_main_HomDR_data}
\end{figure}

\begin{figure}[t]
\centering
\includegraphics[width=0.4\textwidth]{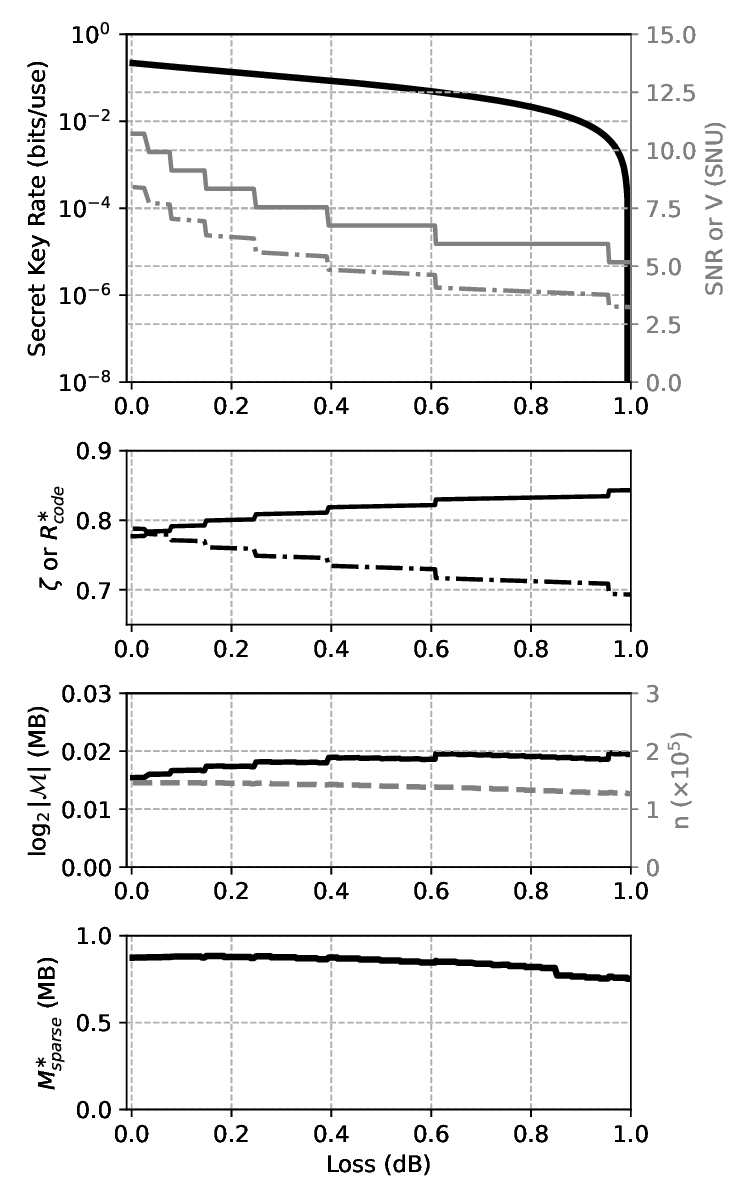}
\caption{In the top panel, we plot the secret key rate of Eq.~\eqref{main result} (black line) of the Gaussian modulation protocol with coherent states and homodyne detection in RR against the loss in dB. The rest of the parameters and settings have been considered the same as in Fig.~ \ref{fig:script_main_HomDR_data}. Here we observe a better performance in terms of tolerable loss compared with the direct reconciliation protocol in Fig.~\ref{fig:script_main_HomDR_data} (see also Fig.~\ref{fig:main_mixed}). This is expected due to the known $3$ dB loss limit of the DR protocol in the ideal asymptotic regime which decreases even more when one considers finite-size effects and channel noise.}
\label{fig:script_main_HomRR_data}
\end{figure}

\begin{figure}[t]
\centering
\includegraphics[width=0.4\textwidth]{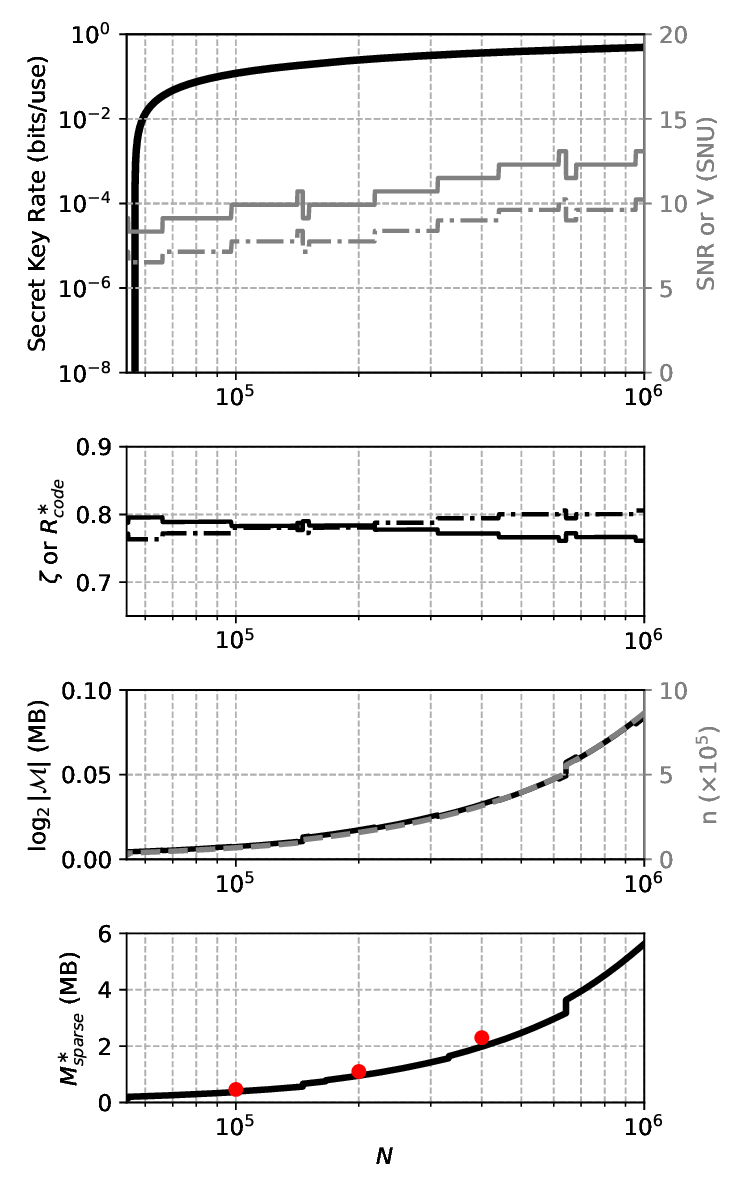}
\caption{In the first panel, we plot the secret key rate of Eq.~\eqref{main result} (black line) for the Gaussian modulation protocol with coherent states and homodyne detection in direct reconciliation against $N$. We set {the loss to $0.02~\text{dB}$}. We have optimized over $V$ and $n/N$. The rest of the parameters, settings and lines are the same as in Fig.~\ref{fig:script_main_HomDR_data}. With red ink, we plot the points included in Table~\ref{table:VALUE_TABLE} of the corresponding $M_\text{sparse}$, i.e., the actual storage needed for these sparse matrices after creating them in CRS format using the GMCS simulation library in~\cite{softquanta2021}.}
\label{fig:script_db=0.02_HomDR_data}
\end{figure}

\begin{figure}[t]
\centering
\includegraphics[width=0.4\textwidth]{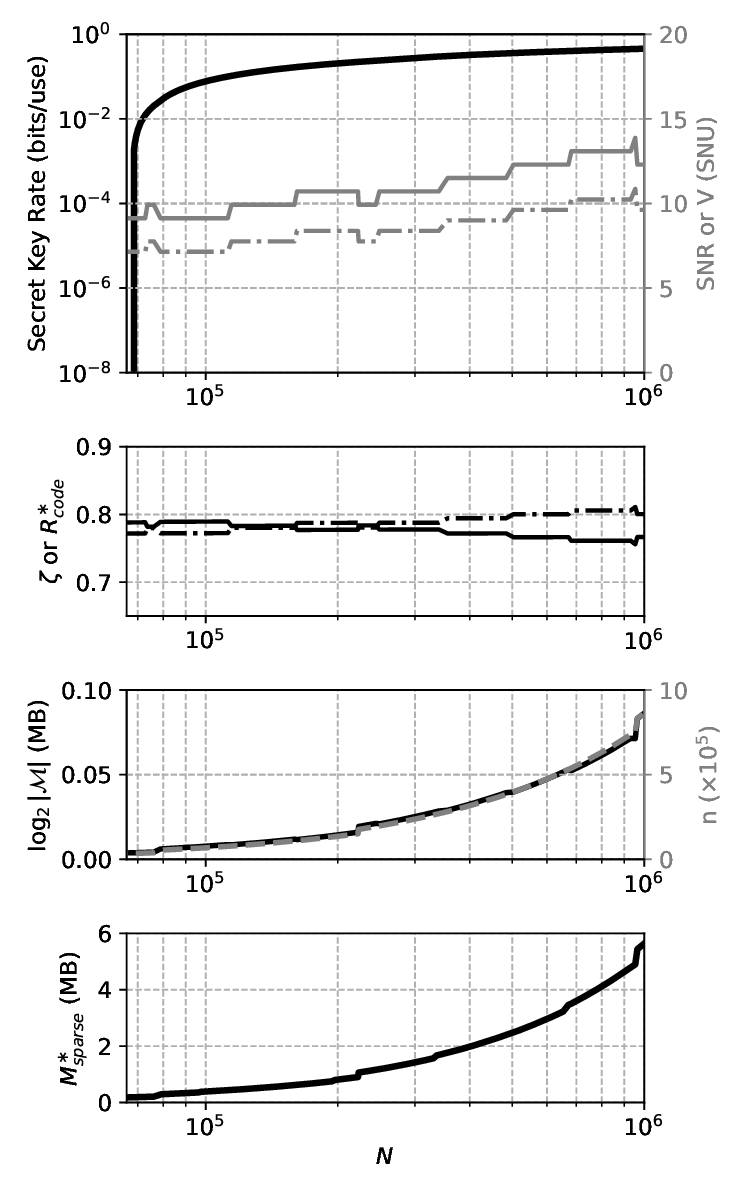}
\caption{In the first panel, we plot the secret key rate of Eq.~\eqref{main result} (black line) for the Gaussian modulation protocol with coherent states and homodyne detection in RR against $N$. We have set the loss to $0.02$ dB. The rest of the parameters, settings and lines are the same as in Fig.~\ref{fig:script_main_HomDR_data}. We observe here, that we need a larger block size in order to obtain a secret key rate compared to the DR protocol in Fig.~\ref{fig:script_db=0.02_HomDR_data} (see also Fig.~\ref{fig:blocksize_mixed}). This is because the DR protocol offers higher rates than the RR protocol in the low loss regime.}
\label{fig:script_db=0.02_HomRR_data}
\end{figure}

\section{Results\label{Results}}
In this section, we connect the previous theoretical results with the practical implications for the protocol operations and especially the data post-processing part. We focus on the protocol with homodyne detection. In Fig.~\ref{fig:script_main_HomDR_data}, we plot the secret key rate of Eq.~\eqref{main result} against the loss in dB, where we have optimized over $V$ and the PE ratio $1-(n/N)$. We also plot the changes in terms of $\zeta$ or leakage and the associated SNR for the optimized rate value. These correspond to the values of $V$ and $n$ presented in the same figure. Furthermore, using the same parameters for every point in the plot of the secret key rate, we have plotted the corresponding points for $M_\text{sparse}$ in Eq.~\eqref{mem_gen_sparse}. The rest of the parameters used for this plot are summarized in Table~\ref{table:comp_rate_params}.

In Fig.~\ref{fig:script_main_HomRR_data}, we plot the corresponding RR secret key rate, where it is clear that the performance has increased in terms of loss tolerance. This is expected due to the 3 dB loss limit of the direct reconciliation protocol in the asymptotic regime, which degrades when one assumes finite-size effects. Clearly, the RR protocol is robust against higher losses (see Fig.~\ref{fig:script_db=2_HomRR_data}) especially when one uses a large block size. This is not achievable by the DR protocol. We also include plots for the corresponding $R^\ast_\text{code}$ that is needed to achieve the specific performance for the given SNR and, similarly, the corresponding $M^\ast_\text{sparse}$. 

By contrast, in Figs.~\ref{fig:script_db=0.02_HomDR_data} and~\ref{fig:script_db=0.02_HomRR_data}, we plot the secret key rates for the DR and RR protocols, respectively, against the block size. We set the loss to $0.02$ dB and choose the other parameters in the same way. We mainly observe that the DR protocol is advantageous in this regime of losses: we can achieve high rates by using smaller bock-size. Here, we consider a moderate block size of the order of $10^5-10^6$. This is because we would like to investigate regimes of operation where the high-rate performance in long distance is not a priority. These regimes are described by fast sharing of small keys assuming the smallest requirements of hardware equipment, either because of space constraints or cost-effective implementations.  For example, this can be described by QKD implementations over networks of small sensor devices, Internet of Things (IoT) nodes, wearable devices, or drones operating inside a building, outside, connected with fibre or with free-space links~\cite{RazaviIndoors,free-space}.

Apart from the limitations due to the communication links such as the noise and the losses, one should take into account as a priority the hardware requirements of the classical data post-processing. This can be done effectively, in the finite-size regime, using composable terms connected to every performance aspect. Therefore, we combine tools developed in previous studies to characterize the requirements in storage during one-way EC and especially the encoding part which is executed by one of the parties.  Note that the DR protocol is advantageous in this regime because it can give higher rates for smaller block size. In fact, it can support lightweight and agile transmitters responsible for the EC encoding while the bulkier receivers can be better suited for computationally intensive tasks such as the EC decoding in an asymmetric scenario: for example, a network of small sensors transmitting to a central processing unit.  

The syndrome creation, i.e., encoding process, is the less difficult part of the one-way EC with LDPC codes. In contrast, the decoding process is rather demanding and can be effectively handled by larger stations rather than a constrained device.  In such an asymmetric scenario in terms of computational power, for an appropriate loss tolerance, the party operating through the constrained device is the transmitter and also the encoder during EC. This describes the DR protocol. In this setting, one can exploit the trade-off between limitations in robustness against losses with the mitigation of computational-power requirements. 

Then, one still needs to check the compatibility with storage requirements for the parity-check matrix as the main aspect of the EC encoding procedure. In particular, the amount of leakage calculated through Eq.~\eqref{tight_bound} that achieves a specific performance in terms of secret key rate in Eqs.~\eqref{main result} can be mapped to the associated LDPC code rate in Eq.~\eqref{Rcode_ast} and, in turn, to the dimensions of the related parity-check matrices that give the associated leakage. Then we can predict the related storage requirements by Eq.~\eqref{mem_gen_sparse_theo}. 

In Fig.~\ref{fig:script_db=0.02_HomDR_data}, we can see the behaviour of $M^\ast_\text{sparse}$ against the block size. For specific block sizes, we have created the parity-check matrices for the corresponding encoding process in EC and stored them in sparse form. The details of these parity-check matrices are presented in Table~\ref{table:VALUE_TABLE} while, for the sake of comparison,  the storage for each matrix is described by the red points in the bottom panel. We see that the results are very close to the predicted values for $M_\text{sparse}$. Finally, in Appendix~\ref{other_res}, we examine the behaviour of the protocols in terms of losses when one considers different values for block size, successful EC probability, and digitization. 

Based on this analysis, short-range networks of devices with a constrained EC encoding memory may support CV-QKD connections. This is a good starting point which with further refinements and support from technological advances can lead to constrained-device networks such as IoT, drones, wearable devices to fully support CV-QKD connections.

\section{Conclusion\label{conc}}
Small device detectors and IoT device networks play a crucial role in modern society, enabling real-time monitoring, automation, and seamless connectivity across various sectors, including healthcare, smart cities, industrial automation, and environmental monitoring. Their ability to collect and process vast amounts of data enhances efficiency, reduces human intervention, and supports decision-making in critical applications. However, as these devices become deeply integrated into daily life, their security is of paramount importance. Cyber threats targeting IoT networks can lead to privacy breaches, unauthorized surveillance, or even large-scale disruptions in infrastructure, posing risks to public safety and economic stability.

 Robust solutions against cyber threats are therefore paramount for these devices: one of the main candidates is QKD offering an information-theoretic security advantage. Since these devices operate at short distances, such as within a room, a house, or a warehouse, CV-QKD which has an advantage in this regime and especially the GMCS protocols can provide higher secret key rates. In particular, by using non-binary LDPC codes in a practical implementation of such a protocol, the parties exploit the high mutual information between their continuous variables. 

However, this is quite challenging due to the increased requirements of the data post-processing stage in computational power or storage, not to mention implementing a QKD protocol on such constrained devices in the first place.   Therefore, in order to investigate the performance under those circumstances, one needs to develop rigorous theoretical tools. Here we have combined a composable security proof taking into account the main stages of the data post-processing of a QKD protocol with a tight bound for EC adapting it to the non-binary LDPC regime. 

This allows us to predict optimal secret key performance in terms of reconciliation efficiency and leakage and match this performance to operational code rates for given signal-to-noise ratios and error correction success probabilities. Based on this, we developed a tool that models the encoding storage requirements for a non-binary LDPC EC associated with the given secret key rate performance.   

We combined a composable security framework for the secret key rate of the GMCS protocols with a finite-size tight bound 
for one-way EC leakage of non-binary LDPC dependent on the given SNR and successful EC probability. Through this tool, one can theoretically calculate the code rate for close to optimum performance and the dimensions of the associated parity-check matrix. In turn, one may calculate the storage requirements of the EC encoding process, which is crucial, for example, for the implementation of CV-QKD with constraint devices. Note here that optimum leakage means optimum value for storage.\\

\section*{CODE AND NUMERICAL IMPLEMENTATION}
The majority of the numerical results and plots in this manuscript were produced using custom Python code developed for the calculation of tight leakage bounds in the context of non-binary LDPC codes. This code represents a central technical contribution of the present work and is publicly available at: \href{https://github.com/eqclabs/tight_bound_leakage.git}{\texttt{eqclabs/tight\_bound\_leakage}}. The three red data points in Fig.~\ref{fig:script_db=0.02_HomDR_data} were obtained using an independent implementation, as referenced in the main text. All simulations were performed on nodes of the \href{https://vikingdocs.york.ac.uk}{Viking High Performance Computing cluster} at the University of York, equipped with a 2-core AMD EPYC3 7643 processor and 12~GB of memory. All repositories are released under the Apache License 2.0 and include documentation to support reproducibility.

\section*{ACKNOWLEDGEMENTS}
P.P. thanks Juan Vieira Giestinhas and Alexander George Mountogiannakis for insightful discussions, and the high performance compute facility, the Viking cluster, of the University of York. This work is supported, in part, by EPSRC and DSIT TMF-uplift: CHEDDAR: Communications Hub For Empowering Distributed ClouD Computing Applications And Research (EP/X040518/l), and (EP/Y037421/l). S.P. acknowledges support from EPSRC and UKRI, via the Integrated Quantum Networks (IQN) Research Hub (EP/Z533208/1).

\begin{figure}[h!]
\centering
\includegraphics[width=0.3\textwidth]{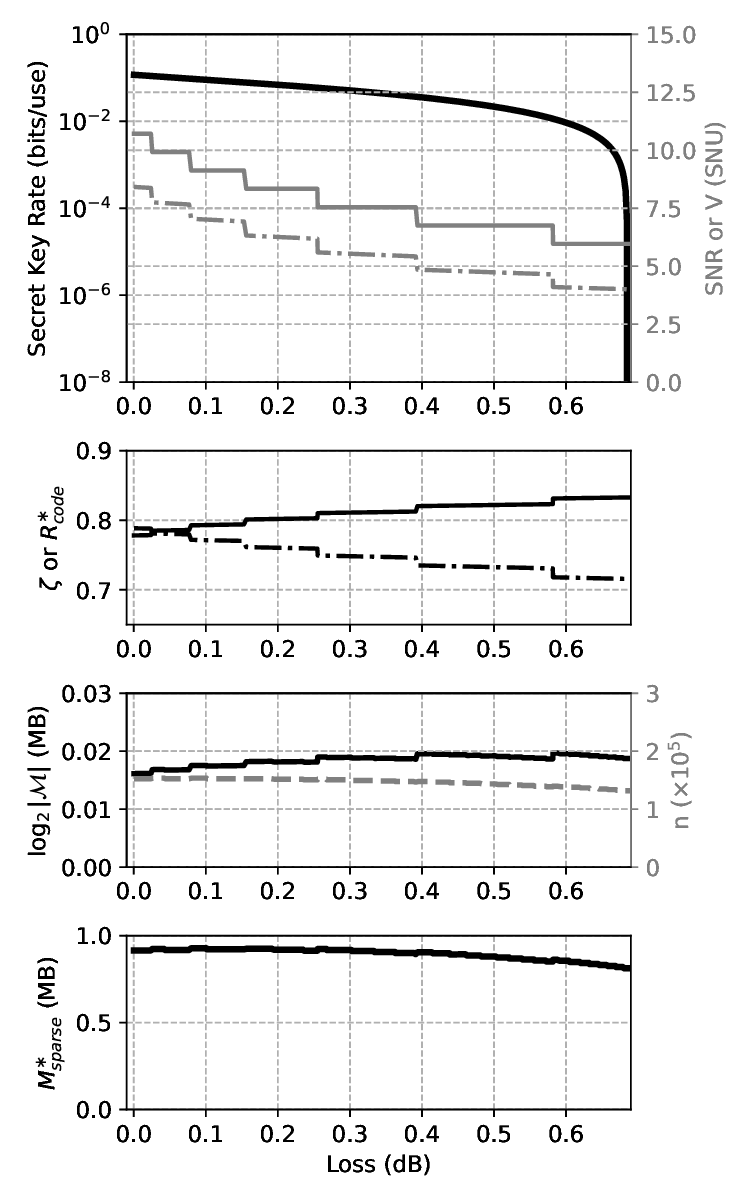}
\caption{We present similar plots to Fig.~\ref{fig:script_main_HomDR_data}. All the parameters are the same, apart from $p_\text{ec}=0.4$. We observe that the secret key rate is lower than the corresponding one in the previous figure. However, the achievable loss has been slightly improved. We see also very similar performance in terms of the other parameters $\zeta$, $R^{\ast}_\text{code}$, leakage, and $M_\text{sparse}$.}
\label{fig:script_pec=0.4_HomDR_data}
\end{figure}

\begin{figure}[h!]
\centering
\includegraphics[width=0.3\textwidth]{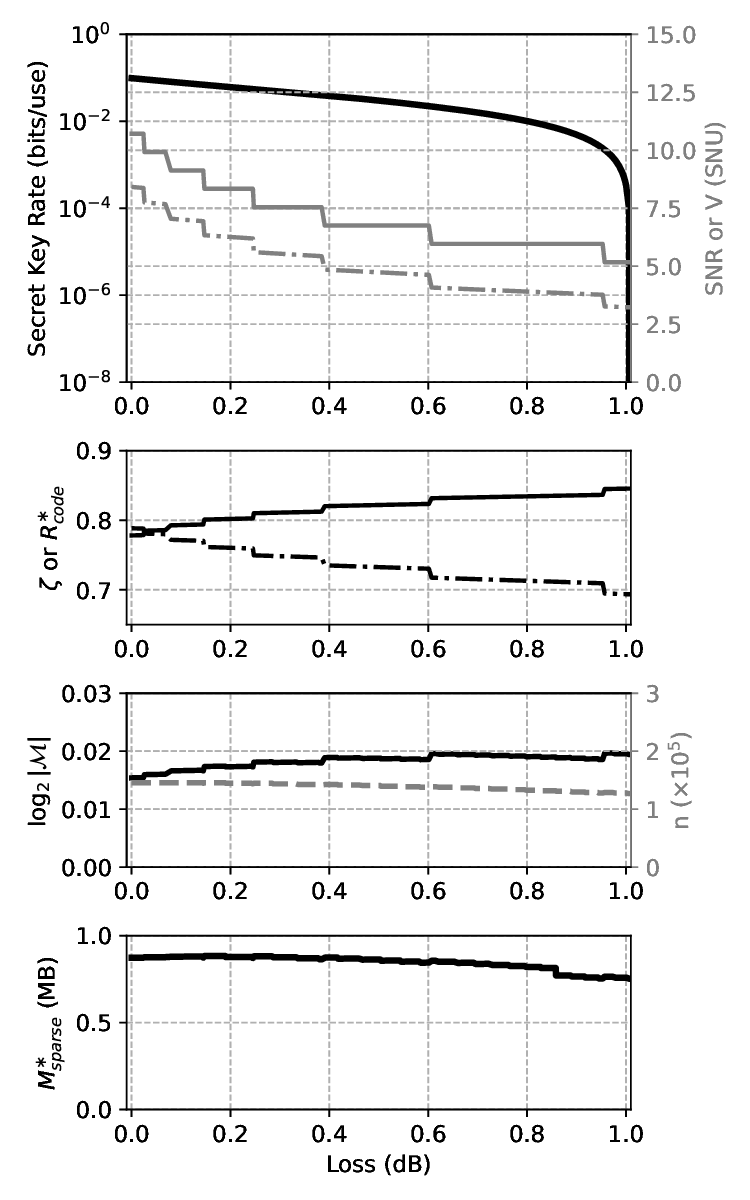}
\caption{We present similar plots to Fig.~\ref{fig:script_main_HomRR_data}. All the parameters are the same, apart from $p_\text{ec}=0.4$. We observe similar behaviour for the rate as in Fig.~\ref{fig:script_pec=0.4_HomDR_data}.}
\label{fig:script_pec=0.4_HomRR_data}
\end{figure}

\begin{figure}[h!]
\centering
\includegraphics[width=0.3\textwidth]{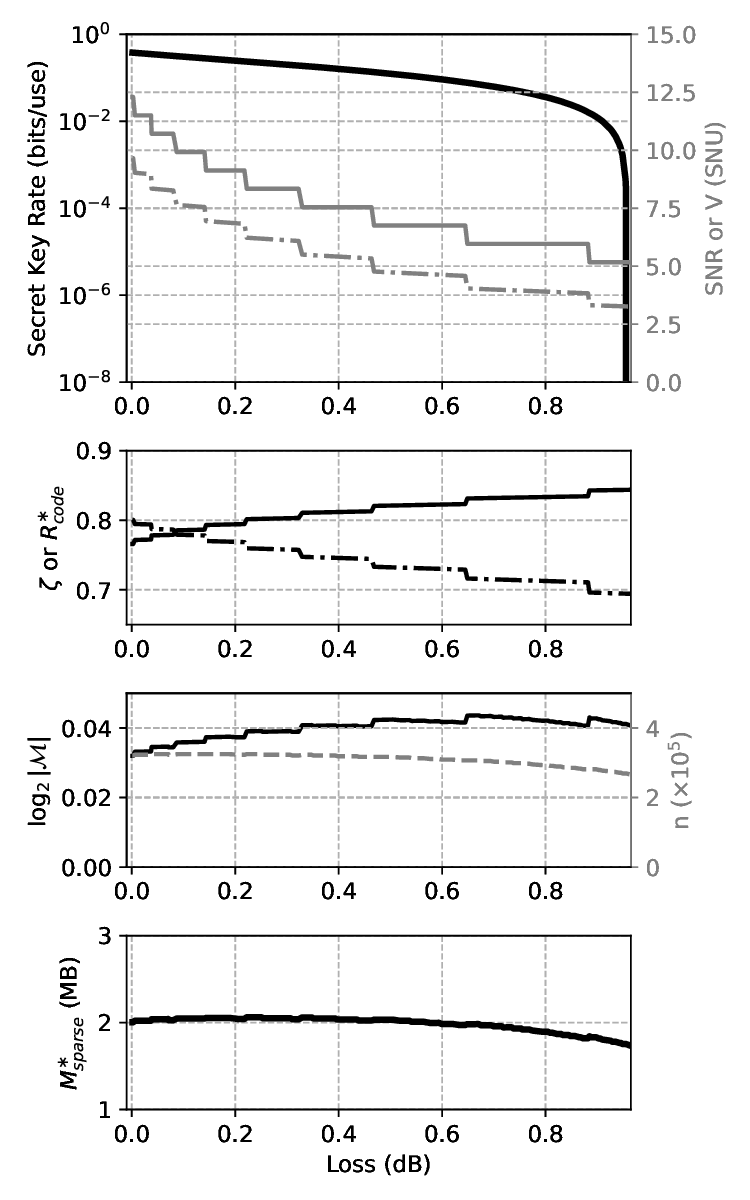}
\caption{We present similar plots  to Fig.~\ref{fig:script_main_HomDR_data}. All the parameters are the same, apart from assuming a doubled value for block size $N=4\times10^5$. We observe that the secret key rate takes higher values and the achievable loss has been significantly increased. The leakage level and $n$ has been increased in the same manner (almost doubled) and similarly $M^\ast_\text{sparse}$. The other parameters have remained in similar levels as before.}
\label{fig:script_bksz=4_HomDR_data}
\end{figure}

\begin{figure}[h!]
\centering
\includegraphics[width=0.3\textwidth]{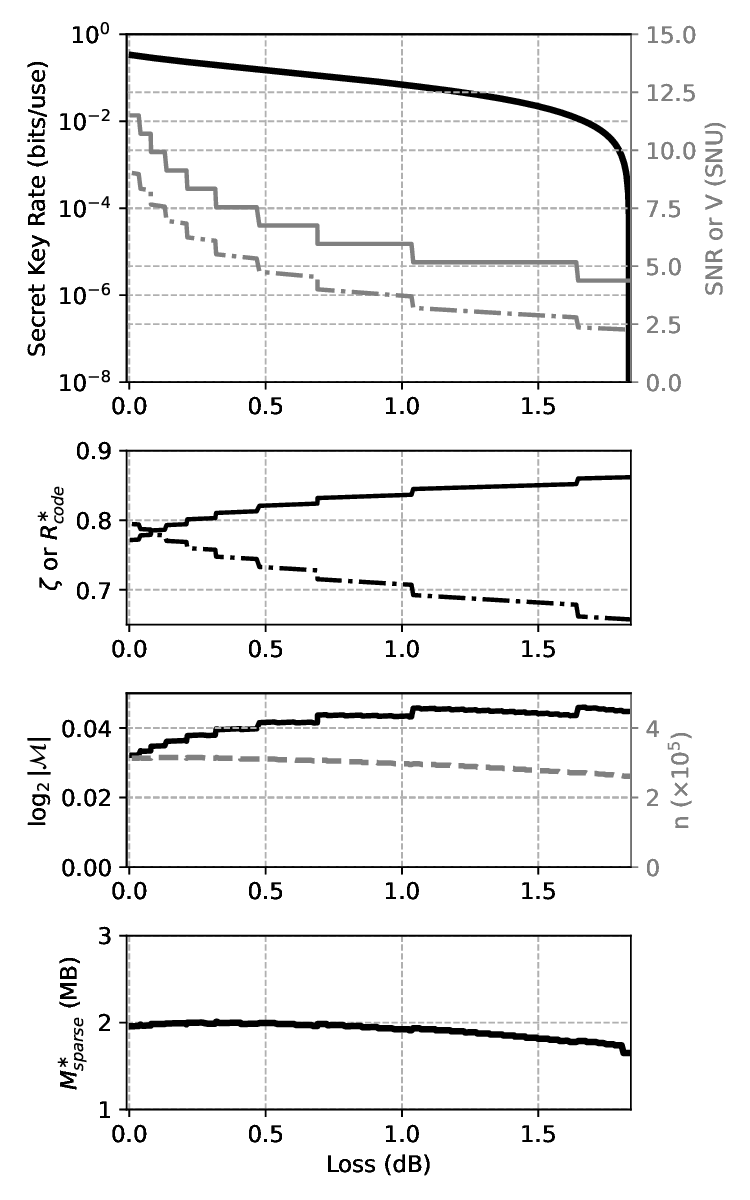}
\caption{We present similar plots to Fig.~\ref{fig:script_main_HomRR_data}. All the parameters are the same, apart from assuming a doubled value for block size $N=4\times10^5$. We observe similar behaviour for the rate as in Fig.~\ref{fig:script_bksz=4_HomDR_data}.}
\label{fig:script_bksz=4_HomRR_data}
\end{figure}

\begin{figure}[h!]
\centering
\includegraphics[width=0.3\textwidth]{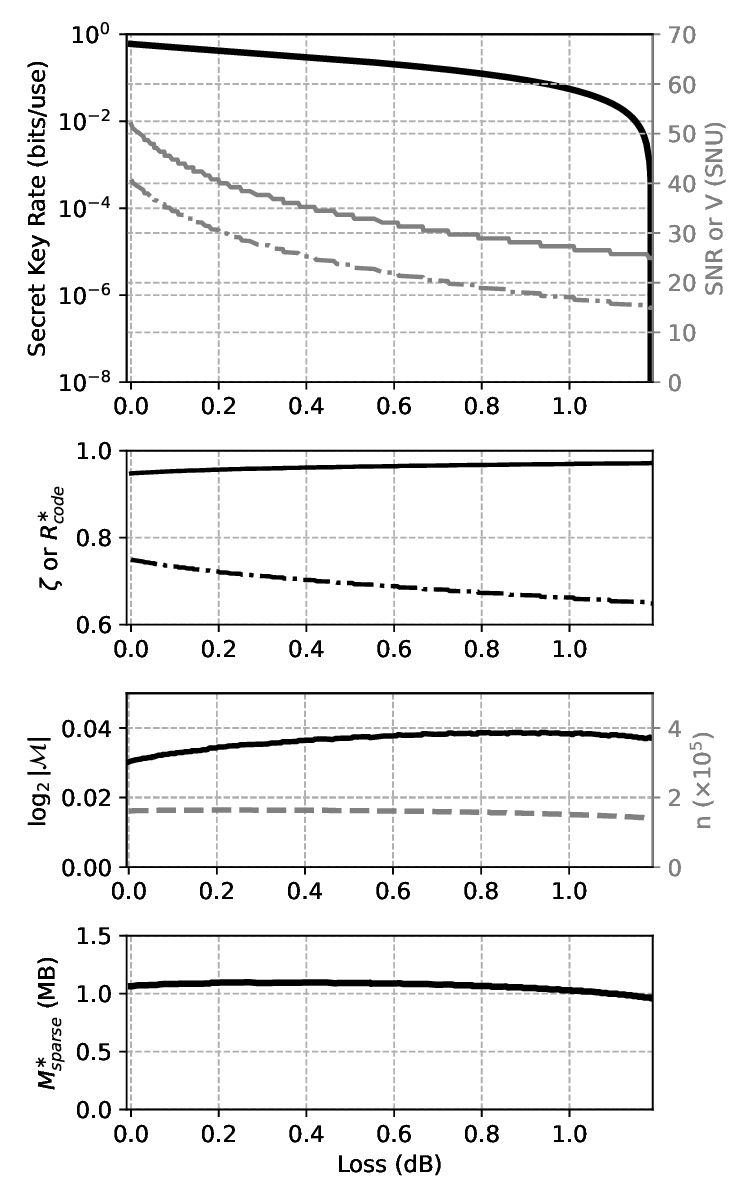}
\caption{We present similar plots to Fig.~\ref{fig:script_main_HomDR_data}. All the parameters are the same, apart from assuming a digitization of $d=6$. We observe that the secret key rate takes higher values and the achievable loss has been significantly increased. The improvement is better compared to increasing the block size as in the case of Fig.~\ref{fig:script_bksz=4_HomDR_data}. By increasing the digitization, we allow the choice for larger optimal values for the Gaussian modulation variance which leads to higher SNRs and very high reconciliation efficiency.}
\label{fig:script_p=6_HomDR_data}
\end{figure}

\begin{figure}[h!]
\centering
\includegraphics[width=0.3\textwidth]{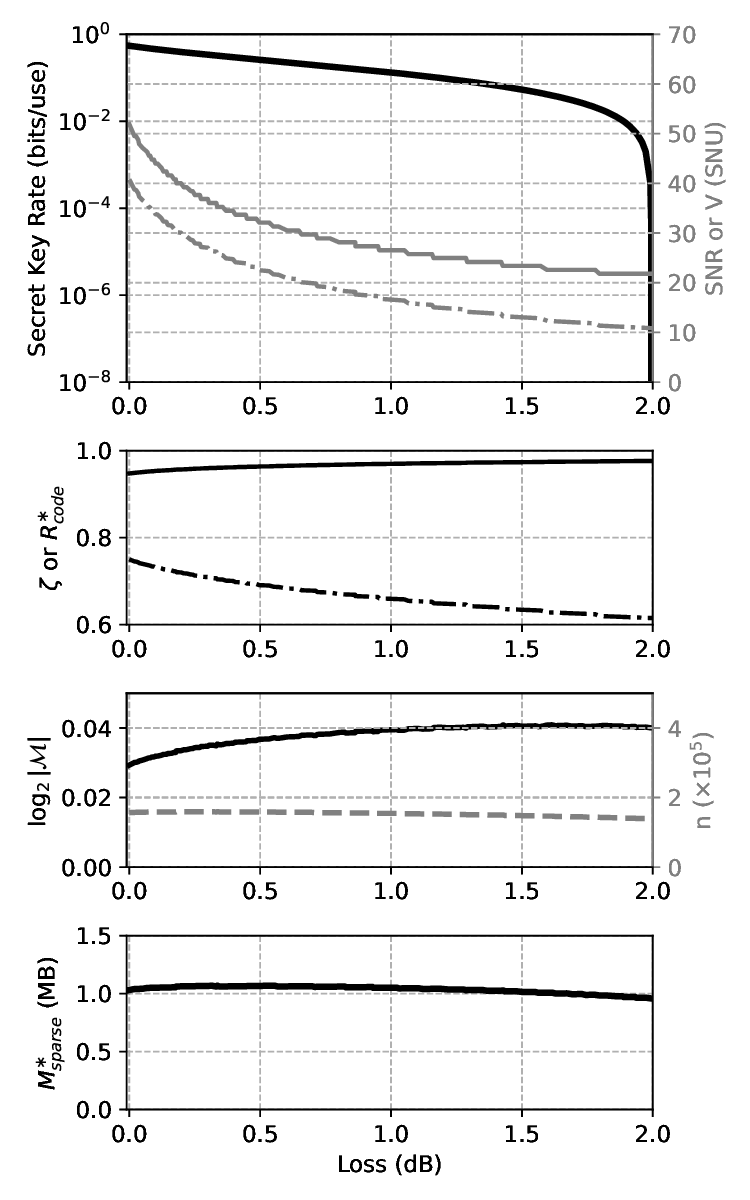}
\caption{We present similar plots to Fig.~\ref{fig:script_main_HomRR_data}. All the parameters are the same, apart from assuming a digitization of $d=6$. We observe similar behaviour for the rate as in Fig.~\ref{fig:script_p=6_HomDR_data}.}
\label{fig:script_p=6_HomRR_data}
\end{figure}

\begin{figure}[h!]
\centering
\includegraphics[width=0.3\textwidth]{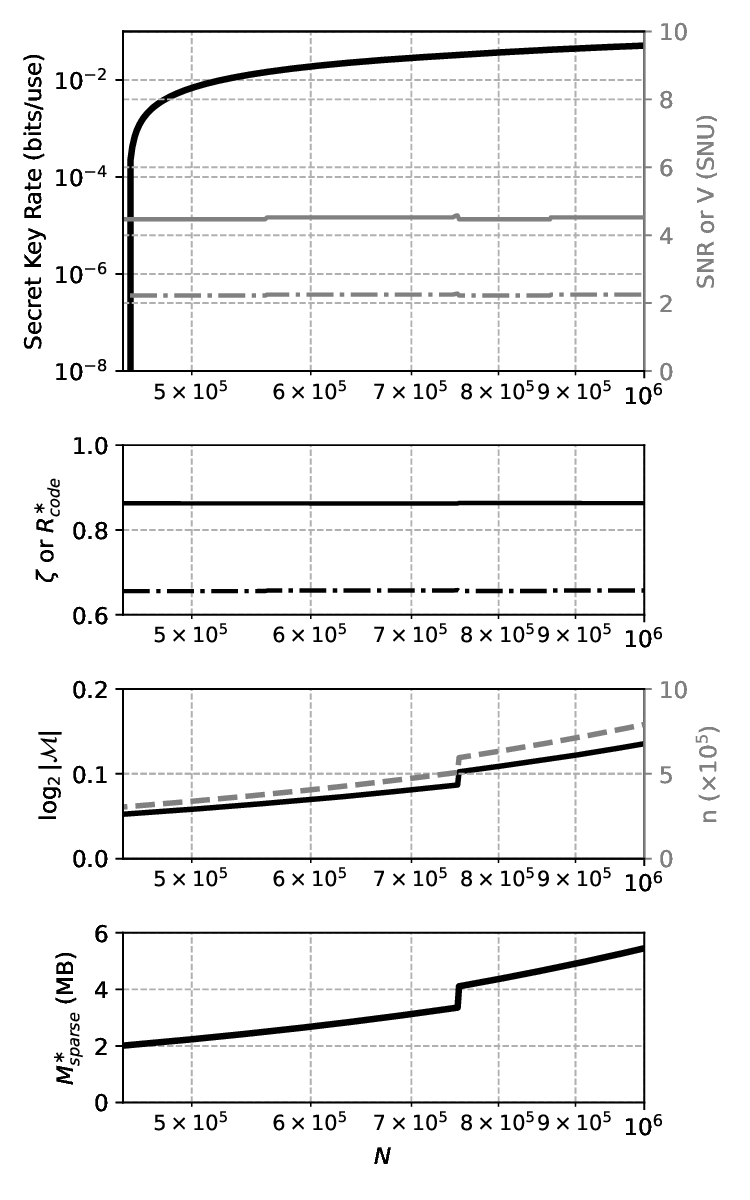}
\caption{We present similar plots to Fig.~\ref{fig:script_db=0.02_HomRR_data}. All the parameters are the same, apart from setting the losses to $2\text{dB}$. We observe that the RR protocol can operate in higher losses given an increased block size above $4 \times 10^5$.}
\label{fig:script_db=2_HomRR_data}
\end{figure}
\begin{figure}[h!]
\includegraphics[width=0.3\textwidth]{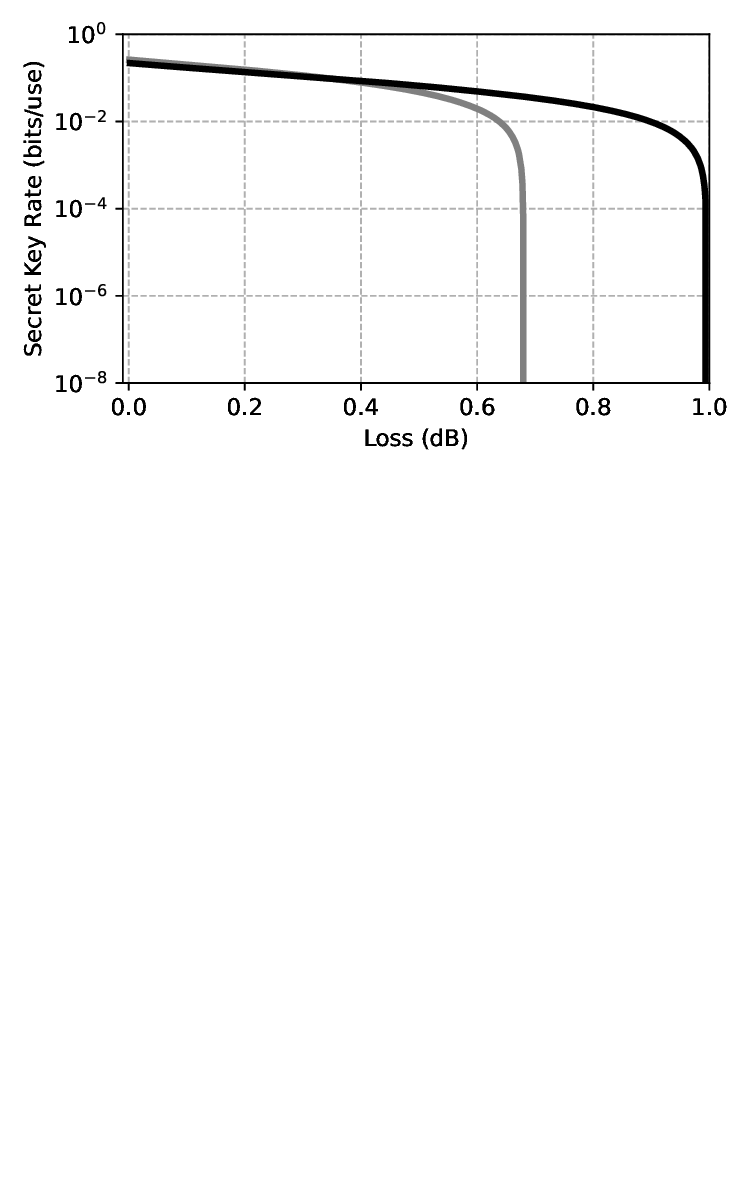}
\caption{We plot the secret key rates from Fig~\ref{fig:script_main_HomDR_data} and~\ref{fig:script_main_HomRR_data} in the same frame for a clear comparison. The RR secret key rate (black line) can surpass the the DR one (gray line) in terms of tolerable losses.}
\label{fig:main_mixed}
\end{figure}
\begin{figure}[h!]
\includegraphics[width=0.3\textwidth]{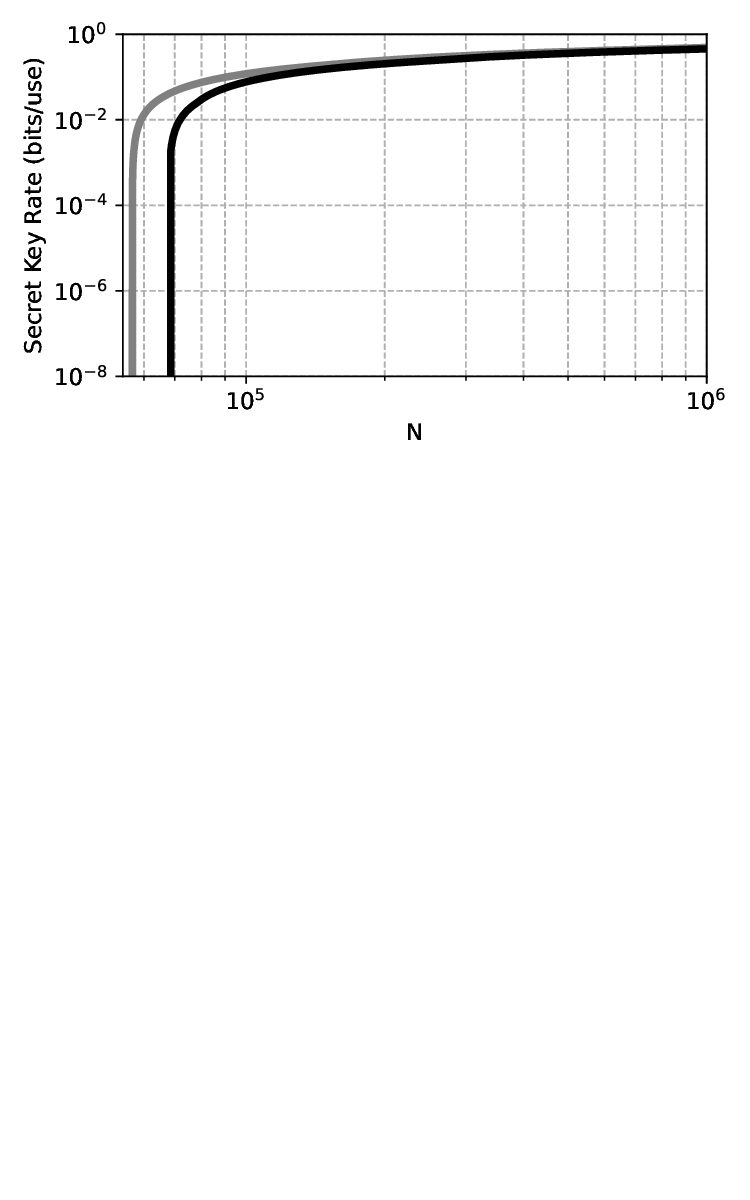}
\caption{We plot the secret key rates from Fig~\ref{fig:script_db=0.02_HomDR_data} and~\ref{fig:script_db=0.02_HomRR_data} in the same frame for a clear comparison. The RR secret key rate (black line) needs a larger block size compared to the DR one (gray line).}
\label{fig:blocksize_mixed}
\end{figure}

\appendix
\section{Virtual concatenation\label{VC}}
Before digitization, the parties may apply classical data post-processing procedures such as rotations and re-scaling which bring the CM in its normal form, i.e., cross-quadrature correlation terms vanish. After digitization, one may apply the concatenation step useful only on the protocols using heterodyne detection. This is a virtual step that results in a common description of the GMCS protocols using either homodyne or heterodyne detection. This is summarised by
\begin{equation}
\mathsf{k}=k
\end{equation}
for the case of homodyne detection and
\begin{equation}\label{vc}
\mathsf{k}=k_q2^d+k_p
\end{equation}
for the case of heterodyne detection. We observe that
\begin{align}
p(\mathsf{k})&=p(k_q,k_p)=p(k_q)p(k_p), ~~k_q=k_p=k\notag\\&\Rightarrow~~p(k_q)=p(k_p).
\end{align}
Therefore,
\begin{align}\label{double_entropy}
H(\mathsf{k})&=H(k_q,k_p)=-\sum_{k_q,k_p}p(k_q,k_p)\log_2p(k_q,k_p)\notag\\
=&-\sum_{k_q,k_p}p(k_q,k_p)\log_2p(k_q)-\sum_{k_q,k_p}p(k_q,k_p)\log_2p(k_p)\notag\\
=&-\sum_{k_q}p(k_q)\log_2p(k_q)-\sum_{k_p}p(k_p)\log_2p(k_p)\notag\\
=&-\sum_{k}p(k)\log_2p(k)-\sum_{k}p(k)\log_2p(k)=2H(k).
\end{align}


\section{Secret key rate Summary \label{sum_SKR}}
Here we write in a more concise way the final equation of the secret key rate used for the calculations of the previous results. By combining Eq.~\eqref{rate_zeta} with Eq.\eqref{secret_key_ration} and multiplying with the term $\frac{n}{N}$ connected with the PE ratio and the probability of remaining blocks $p_\text{ec}$, we obtain
\begin{align}
R:=&p_\text{ec}\frac{n}{N}\bigg[\big( \zeta I(x:y)-\chi(x:E)|_{\tau^{\epsilon_\text{pe}}, \xi^{\epsilon_\text{pe}}} \big)\notag\\
&-\frac{\Delta^{\epsilon_\text{s}}_\text{aep}(h)}{\sqrt{n}}+\frac{\theta}{n} -\mathcal{O}\left(\delta(n)/ n\right)\bigg],
\end{align}
where $\zeta$ is given by Eq.~\eqref{rec_eff_theo}, $\theta$ is given by Eq.~\eqref{theta}, and $\Delta^{\epsilon_\text{s}}_\text{aep}(h)$  by Eq.~\eqref{delta_aep}. For RR, the mutual information $I(x:y)$ and the Holevo information $\chi(x:E)$ are given in Ref.~\cite{ma_sim_one,ma_sim_two}. For DD, one uses the same rational as in Ref.~\cite{ma_sim_one,ma_sim_two} but using the following conditional CMs: for heterodyne detection
\begin{equation}
\mathbf{V}_{E|x}=\begin{pmatrix}
\diag\{\phi_0,\phi\} &\psi\mathbf{Z}\\
\psi\mathbf{Z}&\omega \mathbf{I}
\end{pmatrix}
\end{equation}
 and for homodyne detection
\begin{equation}
\mathbf{V}_{E|x}=\begin{pmatrix}
\phi_0 \mathbf{I} &\psi\mathbf{Z}\\
\psi\mathbf{Z}&\omega \mathbf{I}
\end{pmatrix}
\end{equation}
with
\begin{align}
\phi_0=&\tau \omega+(1-\tau),\\
\phi=&\tau \omega+(1-\tau)(V+1),\\
\psi=&\sqrt{\tau (\omega^2-1)},
\end{align}
where $\omega$ is Eve's noise variance, $V$ is the Gaussian classical modulation, and $\tau$ is the transmissivity of the channel. Note that the latter are not dependent on the detection efficiency and or the electronic noise at Bob's lab.

\section{Other results\label{other_res}}
We now investigate the behaviour of the protocols, in terms of the secret key rate against losses, assuming either a different successful EC probability, block size, or digitization parameter. For example, we see that changing the $p_\text{ec}$ from $0.9$ to $0.4$ affects the performance of both reconciliation directions: in Figs.~\ref{fig:script_pec=0.4_HomDR_data} and~\ref{fig:script_pec=0.4_HomRR_data}, we see a considerable drop in terms of the rate but a minimal improvement in loss tolerance compared to Figs. ~\ref{fig:script_main_HomDR_data} and~\ref{fig:script_main_HomRR_data}, respectively.

In Figs.~\ref{fig:script_bksz=4_HomDR_data} and~\ref{fig:script_bksz=4_HomRR_data}, we doubled the block size to $4\times 10^5$ compared to Figs. ~\ref{fig:script_main_HomDR_data} and~\ref{fig:script_main_HomRR_data},respectively. Here we see an improvement in the performance and loss tolerance: we obtain rates almost in the $3/2$ amount or more of loss in both cases compared to Figs. ~\ref{fig:script_main_HomDR_data} and~\ref{fig:script_main_HomRR_data}, respectively. However, the amount of $M^\ast_\text{sparse}$ required is almost doubled as expected by the linear dependence of $M^\ast_\text{sparse}$ on block size.

When one increases the digitization parameter (here from $d=4$ to $6$), the performance also increases along with the loss tolerance. We show this tendency in Figs.~\ref{fig:script_p=6_HomDR_data} and~\ref{fig:script_p=6_HomRR_data} for both protocols, respectively. In particular, the tolerable loss is larger than the double in Figs. ~\ref{fig:script_main_HomDR_data} and~\ref{fig:script_main_HomRR_data}. In addition, the leakage and $M^\ast_\text{sparse}$ increase slightly.
This means that by increasing the digitization parameter, we can obtain similar performance as by increasing the block size avoiding large storage requirements.

Increasing the digitization means that $\zeta_\text{digit}$ approaches $1$. In other words, the parties can exploit almost the whole amount of the CV mutual information available to them. This is why CV are advantageous especially for small loss. However, the terms $\Delta^{\epsilon_\text{s}}_\text{aep}$ and $\Delta^{\epsilon_\text{ec}}_\text{leak}$ (also expressed through $\zeta_\text{leak}$) increase with larger digitization parameters. This will lead to a saturation point for the secret key rate performance and loss tolerance. 

 In Fig.~\ref{fig:script_db=2_HomRR_data}, we plot the secret key rate for the RR protocol  against the block size for $\mathrm{dB}=2$. We observe that the RR protocol can tolerate higher losses. Finally, in Fig.~\ref{fig:main_mixed}, we have plotted together the secret key rate plots from Figs.~\ref{fig:script_main_HomDR_data} and~\ref{fig:script_main_HomRR_data} for a clearer comparison. We have done the same for Figs.~\ref{fig:script_db=0.02_HomDR_data} and ~\ref{fig:script_db=0.02_HomRR_data} in Fig.~\ref{fig:blocksize_mixed}.

\end{document}